\newcommand{\be}{\begin{eqnarray}}
\newcommand{\ee}{\end{eqnarray}}
\documentstyle[aps,epsfig]{revtex}

\begin{document}
%\tolerance=10000
%\newcommand{\be}{\begin{eqnarray}}
%\newcommand{\ee}{\end{eqnarray}}
%\documentstyle[preprint,aps,epsfig]{revtex}
%\unitlength=1mm
%\begin{document}
%\tightenlines
\draft
\title{Pion propagation in the linear sigma model at finite temperature}
\author{Alejandro Ayala, Sarira Sahu}
\address{Instituto de Ciencias Nucleares\\
         Universidad Nacional Aut\'onoma de M\'exico\\
         Aptartado Postal 70-543, M\'exico Distrito Federal 04510, M\'exico.}
\maketitle
\begin{abstract}

We construct effective one-loop vertices and propagators in the linear sigma 
model at finite temperature, satisfying the chiral Ward identities and thus 
respecting chiral symmetry, treating the pion momentum, pion mass and 
temperature as small compared to the sigma mass. We use these objects to 
compute the two-loop pion self-energy. We find that the perturbative behavior 
of physical quantities, such as the temperature dependence of the pion mass, 
is well defined in this kinematical regime in terms of the parameter 
$m_\pi^2/4\pi^2f_\pi^2$ and show that an expansion in terms of this reproduces 
the dispersion curve obtained by means of chiral perturbation theory at 
leading order. The temperature dependence of the pion mass is such that the 
first and second order corrections in the above parameter have the same sign. 
We also study pion damping both in the elastic and inelastic channels to this 
order and compute the mean free path and mean collision time for a pion 
traveling in the medium before forming a sigma resonance and find a very good 
agreement with the result from chiral perturbation theory when using a value 
for the sigma mass of $600$ MeV.

\end{abstract}

\section{Introduction}

A basic ingredient for the understanding of several physical processes 
taking place in hadronic plasmas are the propagation properties of pions. The
relevant scenarios include dense astrophysical objects such as the cores of
neutron stars and the evolution of the highly interacting region formed 
in the midsts of high-energy heavy-ion collisions. In the latter 
context, pions are the most copiously produced particles and their (mainly
attractive) interactions within a dense nuclear environment have been 
speculated to give rise to interesting collective surface 
phenomena~\cite{Shuryak}. Moreover, it is well known that the pion group 
velocity is an important piece of information necessary to properly account 
for the component of the dilepton spectrum originated in the hadronic phase, 
below the critical temperatures and densities for the formation of a 
quark-gluon plasma~\cite{Gale-Liu}.

The hadronic degrees of freedom are usually accounted for by means of 
effective chiral theories whose basic ingredient is the fact that pions are
Goldston bosons coming from the spontaneous breakdown of chiral symmetry.
Chiral perturbation theory (ChPT) is one of such effective theories that 
has been employed to show the well known result that at leading 
perturbative order and at low momentum, the modification of the pion
dispersion curve in a pion medium is just a constant, temperature dependent,
increase of the pion mass~\cite{Gasser}. In spite of the mounting complexity
introduced by the number of couplings required at next to
leading order, ChPT has also been used in a two-loop computation of the pion 
self-energy~\cite{Schenk} and decay constant~\cite{Toublan}. A striking result 
obtained from such computations is that at second order, the shift in the 
temperature dependence of the pion mass is opposite in sign and about three
times larger in magnitude than the first order shift, already at temperatures 
on the order of $150$ MeV. This result might signal either the breakdown of 
the perturbative scheme at such temperatures or the need to compute beyond 
next to leading order in ChPT, given the large relative corrections found.  

Nevertheless, the simplest realization of chiral symmetry is still provided by 
the much studied linear sigma model which possesses the convenient feature of
being a renormalizable field theory, both at zero~\cite{Lee} and
(consequently) at finite temperature~\cite{Mohan}. A further motivation to
study this model stems from recent theoretical results ~\cite{Tornqvist} and 
analyses of data~\cite{Svec} that seem to confirm, though not without 
controversy~\cite{Isgur-Harada}, that a broad scalar resonance, with a mass 
in the vicinity of 600 MeV --that can be identified with the $\sigma$-meson-- 
indeed exists. 

In addition to the above mentioned characteristics, the linear sigma model 
also reproduces~\cite{Ayala} the leading order modification to the pion
mass in a thermal pion medium, when use is made of a systematic expansion in 
the ratio $m_\pi^2/m_\sigma^2$ at zeroth order, where $m_\sigma$, $m_\pi$ are 
the vacuum sigma and pion masses, respectively, and when treating the 
momentum and temperature as small quantities of order $m_\pi$. In this regime, 
the effective expansion parameter becomes $m_\pi^2/4\pi^2f_\pi^2$, where
$f_\pi$ is the pion decay constant. The existence of such parameter allows for 
a controlled loop expansion of the pion self-energy from where we can extract,
for example, the modification to the pion mass at next to leading order.

In this work, we use the linear sigma model to compute the two-loop
order modification to the pion propagator in a pion medium. 
To this end, we construct one-loop effective vertices and propagators 
satisfying the chiral Ward identities. The net result is a next to leading 
order calculation in the parameter $m_\pi^2/4\pi^2f_\pi^2$, and to zeroth
order in the parameter $m_\pi^2/m_\sigma^2$. From the effective pion 
propagator, we explore the behavior of the pion dispersion curve at small 
momentum and for temperatures on the order of the pion mass. We also study 
pion damping, both in the inelastic and elastic channels. 

The work is organized as follows: In section~\ref{secII}, we construct an 
effective one-loop sigma propagator and one-sigma two-pion and four-pion 
vertices, working at zeroth order in the parameter $m_\pi^2/m_\sigma^2$ and 
satisfying the chiral Ward identities. In section~\ref{secIII}, we use these 
effective vertices and propagator to compute the pion self-energy at two-loop 
order where the effective expansion parameter turns out to be 
$m_\pi^2/4\pi^2f_\pi^2$. We find the pion dispersion curve and compute also 
the thermal modification of the pion mass at this order. In
section~\ref{secIV}, we look at the damping of pions traveling in the pion 
medium. The main contribution to the damping mechanism comes from the 
formation of sigma resonances. We compute the mean free path and mean 
collision time for a medium pion. We also look at the damping mechanism 
originated in the pion elastic scattering processes. We finally summarize our 
results and conclude in section~\ref{secV}. A short appendix deals with the 
renormalization of the sigma model at finite temperature at two-loop order. 

\section{Effective one-loop vertices and chiral Ward identities}\label{secII}

The Lagrangian for the linear sigma model, including only the meson degrees of 
freedom and after the explicit inclusion of the chiral symmetry breaking term, 
can be written as~\cite{Lee}
\be
   {\mathcal{L}}=\frac{1}{2}\left[(\partial{\mathbf{\pi}})^2 +
                (\partial\sigma)^2 - m_\pi^2{\mathbf{\pi}}^2 - 
                m_\sigma^2\sigma^2\right] 
                - \lambda^2 f_\pi\sigma (\sigma^2 + {\mathbf{\pi}}^2) -
                \frac{\lambda^2}{4}(\sigma^2 + {\mathbf{\pi}}^2)^2\, ,
   \label{lagrangian}
\ee
where $\mathbf{\pi}$ and  $\sigma$ are the pion and sigma fields,
respectively, and the coupling $\lambda^2$ is given by
\be
   \lambda^2=\frac{m_\sigma^2-m_\pi^2}{2f_\pi^2}\, .
   \label{coupling}
\ee

From the above Lagrangian one obtains the Green's functions and the Feynman 
rules to be used in perturbative calculations, in the usual manner. In 
particular, the bare pion and sigma propagators 
$\Delta_\pi (P)$, $\Delta_\sigma (Q)$ and the bare one-sigma two-pion and 
four-pion vertices $\Gamma_{12}^{ij}$, $\Gamma_{04}^{ijkl}$ are given by 
(hereafter, capital Roman letters are used to denote four momenta)
\be
   i\Delta_\pi(P)\delta^{ij}&=&\frac{i}{P^2-m_\pi^2}\delta^{ij}\nonumber \\
   i\Delta_\sigma (Q)&=&\frac{i}{Q^2-m_\sigma^2}\nonumber \\
   i\Gamma_{12}^{ij}&=&-2i\lambda^2 f_\pi\delta^{ij}\nonumber \\
   i\Gamma_{04}^{ijkl}&=&-2i\lambda^2(\delta^{ij}\delta^{kl} + 
                          \delta^{ik}\delta^{jl} + \delta^{il}\delta^{jk})\, .
   \label{rules}
\ee
These Green's functions are sufficient to obtain the modification to the 
pion propagator, both at zero and finite temperature, at any given 
perturbative order. 

Alternatively, it is also possible to exploit the relations that chiral 
symmetry imposes among different n-point Green's functions. These
relations, better known as chiral Ward identities (ChWI), are a direct
consequence of the fact that the divergence of the axial current may be used
as an interpolating field for the pion~\cite{Lee}. Thus, one could construct
the modification to one of the above Green's functions at a given perturbative
order and from there, build up the induced modification to other Green's
functions related to the former by a ChWI. For example, two of the ChWI
satisfied --order by order in perturbation theory-- by the functions 
$\Delta_\pi (P)$, $\Delta_\sigma (Q)$, $\Gamma_{12}^{ij}$ and 
$\Gamma_{04}^{ijkl}$ are
\be
   f_\pi\Gamma_{04}^{ijkl}(;0,P_1,P_2,P_3)&=&
   \Gamma_{12}^{kl}(P_1;P_2,P_3)\delta^{ij} + 
   \Gamma_{12}^{lj}(P_2;P_3,P_1)\delta^{ik} +
   \Gamma_{12}^{jk}(P_3;P_1,P_2)\delta^{il}\nonumber \\
   f_\pi\Gamma_{12}^{ij}(Q;0,P)&=&
   \left[\Delta_\sigma^{-1}(Q) - 
   \Delta_\pi^{-1}(P)\right]\delta^{ij}\, ,
   \label{Ward}
\ee
where momentum conservation at the vertices is implied, that is
$P_1+P_2+P_3=0$ and $Q+P=0$. The functional dependence of the vertices in 
Eqs.~(\ref{Ward}) is such that the variables before and after the semicolon 
refer to the four-momenta of the sigma and pion fields, 
respectively~\cite{Lee}.  

At one loop and after renormalization, we recall that the sigma propagator is 
modified by finite terms. At zero temperature this modification is purely 
imaginary and its physical origin is that a sigma particle, with a mass larger 
than twice the mass of the pion, is unstable and has a (large) non-vanishing 
width coming from its decay channel into two pions. At finite temperature the 
modification results in real and imaginary parts. The real part modifies the 
sigma dispersion curve whereas the imaginary part represents a temperature 
dependent contribution to the sigma width.

The one-loop diagrams contributing to the sigma self-energy are depicted in
Fig.~1. Their explicit expressions are
\be
   \Pi^a_\sigma (Q) &=& 6\lambda^4f_\pi^2{\mathcal I}(Q;m_\pi,m_\pi)\, ,
   \nonumber\\
   \Pi^b_\sigma (Q) &=& 18\lambda^4f_\pi^2{\mathcal I}(Q;m_\sigma,m_\sigma)
   \, ,\nonumber\\
   \Pi^c_\sigma &=& 3\lambda^2 {\mathcal J}(;m_\pi)\, ,\nonumber\\
   \Pi^d_\sigma &=& 3\lambda^2 {\mathcal J}(;m_\sigma)\, ,\nonumber\\
   \Pi^e_\sigma &=& -18\frac{\lambda^4}{m_\sigma^2}f_\pi^2 
   {\mathcal J}(;m_\pi)\, ,\nonumber\\
   \Pi^f_\sigma &=& -18\frac{\lambda^4}{m_\sigma^2}f_\pi^2 
   {\mathcal J}(;m_\sigma)\, ,
   \label{sigself}
\ee
where in the imaginary-time formalism of thermal field theory (TFT), the 
functions ${\mathcal I}$ and ${\mathcal J}$ are defined by
\be
   {\mathcal I}(Q;m_i,m_j)&\equiv& T\sum_n\int\frac{d^3k}{(2\pi)^3}
                  \frac{1}{K^2+m_i^2}
                  \,\frac{1}{(K-Q)^2+m_j^2}
                  \, ,\nonumber\\
   {\mathcal J}(;m_i) &\equiv& T\sum_n\int\frac{d^3k}{(2\pi)^3}
          \frac{1}{K^2+m_i^2}
   \label{funcIJ}
\ee
and depend parametrically on the mass $m_i=m_\pi ,\, m_\sigma$. Here
$Q=(\omega,{\mathbf{q}}),\, K=(\omega_n,{\mathbf{k}})$ with $\omega = 2m\pi T$ 
and $\omega_n = 2n\pi T$ ($m$, $n$ integers) being discrete boson frequencies,
T is the temperature and $q=|{\mathbf{q}}|,\, k=|{\mathbf{k}}|$.

%%%%%%%%%%%%%%%%%%%%%%%%%%%%%%%%%%%%%%%%%%%%%%%%%%%%%%%%
\begin{figure}[h!] % fig 1
%\hspace{-2.5cm}
\centerline{\epsfig{file=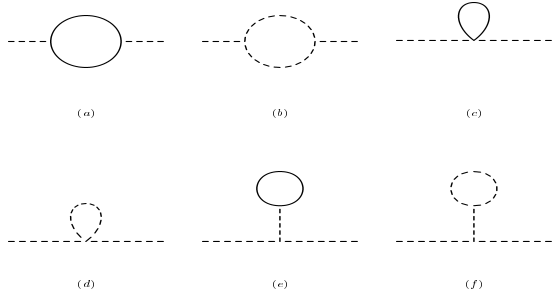,height=1.5in,width=2.5in}}
\vspace{1cm}
\caption{
Diagrams contributing to the sigma self-energy at one loop.
}
\end{figure}
%%%%%%%%%%%%%%%%%%%%%%%%%%%%%%%%%%%%%%%%%%%%%%%%%%%%%%%%

Notice that the functions $\Pi^b_\sigma$ to $\Pi^f_\sigma$ {\it do not} develop
imaginary parts. For $\Pi^b_\sigma$ the several processes described by
a possible imaginary part are kinematically forbidden whereas for 
$\Pi^c_\sigma$ to $\Pi^f_\sigma$ the absence of imaginary parts can be seen
by invoking Cutkosky's rules. Moreover, as it will shortly become clear in
Section~\ref{secIII}, for the real part of ${\mathcal I}$ we require only to 
consider its value for $Q=0$ after the analytical continuation 
\be
   i\omega\rightarrow q_0+i\varepsilon(q_0) q_0\, ,
   \label{continuation}
\ee
which yields the time-ordered version ${\mathcal I}^t$ of the function
${\mathcal I}$ and where $\varepsilon(q_0)$ is the sign function. 
In this limit we have
\be
   {\mbox R}{\mbox e}{\mathcal I}^t(0;m_i,m_i)=-\frac{1}{8\pi^2}\int_0^\infty
   \frac{dx}{\sqrt{1+x^2}}\left[1+2f\left(m_i\sqrt{1+x^2}\right)
   \right]\, ,\
   \label{refunc1}
\ee
whereas ${\mathcal J}$ is given by
\be
   {\mathcal J}(;m_i)=\frac{m_i^2}{4\pi^2}\int_0^\infty
   \frac{dx\, x^2}{\sqrt{1+x^2}}\left[
   1+2f\left(m_i\sqrt{1+x^2}\right)\right]\, 
   \label{refunc2}
\ee
where the function $f$ is the Bose-Einstein distribution
\be
   f(x)=\frac{1}{e^{x/T}-1}\, .
   \label{bose}
\ee
The integrals in Eqs.~(\ref{refunc1}) and~(\ref{refunc2}) contain the usual 
$T=0$ infinities which are taken care of by the introduction of suitable 
counterterms through the renormalization procedure. The temperature-dependent 
terms are expressed as dimensionless functions of the ratio $m_i/T$. These are 
explicitly
\be
   {\mbox R}{\mbox e}{\mathcal I}^t(0;m_i,m_i)&\rightarrow&
   -\frac{1}{4\pi^2}\int_0^\infty
   \frac{dx}{\sqrt{1+x^2}}f\left(m_i\sqrt{1+x^2}\right)\nonumber\\
   &\equiv&-\frac{1}{4\pi^2}h(m_i/T)\, ,\nonumber\\
   {\mathcal J}(;m_i)&\rightarrow&
   \frac{m_i^2}{2\pi^2}\int_0^\infty
   \frac{dx\, x^2}{\sqrt{1+x^2}}
   f\left(m_i\sqrt{1+x^2}\right)\nonumber\\
   &\equiv&\frac{m_i^2}{2\pi^2}g(m_i/T)\, ,
   \label{temprefuncs}
\ee
where the arrows indicate only the temperature dependence of the expressions. 
Notice that for $T\sim m_\pi$, $g(m_\pi/T),\, h(m_\pi/T)\sim 1$. However, for 
$m_i=m_\sigma\sim 600$ MeV, $g$ and $h$ are exponentially small and thus, in
the kinematical regime of interest, we neglect these contributions. The
remaining terms in Eq.~(\ref{sigself}) are just $\Pi_\sigma^a$ and
$\Pi_\sigma^c$. Adding these up, the real part of the sigma self-energy at
$Q=(q_0,{\mathbf{q}})=0$ becomes
\be
   {\mbox R}{\mbox e}\Pi_\sigma (0) = -\frac{6\lambda^4 f_\pi^2}{4\pi^2}
   \left\{ h(m_\pi/T) - \frac{m_\pi^2}{\lambda^2 f_\pi^2} g(m_\pi/T) \right\}
   \, .\label{reap}
\ee
At this point, we make a further simplification. We notice that the ratio
$m_\pi^2/\lambda^2 f_\pi^2$ can be expanded as
\be
   \frac{m_\pi^2}{\lambda^2 f_\pi^2}\sim 2\frac{m_\pi^2}{m_\sigma^2}
   \left( 1 + \frac{m_\pi^2}{m_\sigma^2} + \dots \right)\, ,
   \label{expan}
\ee
and that the leading term in the above expansion is of order
$m_\pi^2/m_\sigma^2$. We will ignore corrections of this order, considering 
the sigma particle as very heavy compared to the pion. In this approximation, 
Eq.~(\ref{reap}) becomes
\be
   {\mbox R}{\mbox e}\Pi_\sigma (0)&\sim&6\lambda^4 f_\pi^2 
   {\mbox R}{\mbox e}{\mathcal I}^t(0;m_\pi,m_\pi)
   \nonumber\\
   &=& -\frac{6\lambda^4 f_\pi^2}{4\pi^2}
   h(m_\pi/T)\, .
   \label{reap2}
\ee
On the other hand, the imaginary part of $\Pi_\sigma$ is expressed in terms of 
the imaginary part of the function ${\mathcal I}^t$, where this is given 
by
\be
   {\mbox I}{\mbox m}{\mathcal I}^t(Q;m_\pi,m_\pi)&=&
   \frac{\varepsilon (q_o)}{2i}\left[
   {\mathcal I}(i\omega\rightarrow q_o + i\epsilon ,q) - 
   {\mathcal I}(i\omega\rightarrow q_o - i\epsilon ,q)\right]
   \nonumber \\
   &=&-\frac{1}{16\pi}
   \left\{ a(Q^2)+\frac{2T}{q}\ln\left(\frac{1-e^{-\omega_+(q_0,q) /T}}
   {1-e^{-\omega_-(q_0,q) /T}}\right)\right\}
   \Theta(Q^2-4m_\pi^2)\, ,
   \label{imtimeorder}
\ee
where $Q^2=q_0^2-q^2$ and $\Theta$ is the step function and the functions 
$a$ and $\omega_\pm$ are 
\be
   a(Q^2)&=&\sqrt{1-\frac{4m_\pi^2}{Q^2}}\nonumber \\
   \omega_\pm (q_0,q)&=&\frac{|q_0| \pm a(Q^2)q}{2}\, .
   \label{imfunc}
\ee
%%%%%%%%%%%%%%%%%%%%%%%%%%%%%%%%%%%%%%%%%%%%%%%%%%%%%%%%
\begin{figure}[h!] % fig 2
%\hspace{-2.5cm}
\centerline{\epsfig{file=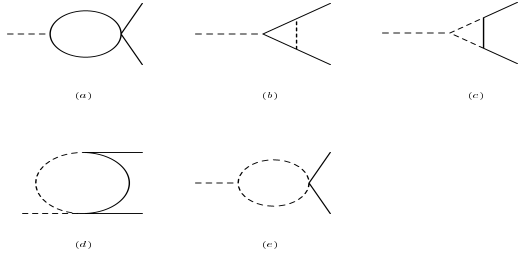,height=1.3in,width=2.5in}}
\vspace{1cm}
\caption{
 Diagrams contributing to the one-sigma two-pion vertex at one loop.
}
\end{figure}
%%%%%%%%%%%%%%%%%%%%%%%%%%%%%%%%%%%%%%%%%%%%%%%%%%%%%%%%
Therefore, the one-loop effective sigma propagator becomes
\be
   i\Delta_\sigma^\star (Q)&=&\frac{i}{Q^2 - m_\sigma^2 + 
                              6\lambda^4f_\pi^2{\mathcal I}^t
                              (Q;m_\pi,m_\pi) }\, .
   \label{newsigprop}
\ee

We now turn to the effective one-sigma two-pion vertex. The diagrams 
contributing at one loop order are shown in Fig.~2, their explicit expressions 
are
\be
   \delta\Gamma_{12}^{a\,\, ij}(Q)&=& 10\lambda^4f_\pi\delta^{ij} 
   {\mathcal I}(Q;m_\pi,m_\pi)\, ,\nonumber\\
   \delta\Gamma_{12}^{b\,\, ij}(Q,P)&=& 8\lambda^6f_\pi^3\delta^{ij} 
   {\mathcal K}(Q,P;m_\pi,m_\sigma)\, ,\nonumber\\
   \delta\Gamma_{12}^{c\,\, ij}(Q,P)&=& 24\lambda^6f_\pi^3\delta^{ij} 
   {\mathcal K}(Q,P;m_\sigma,m_\pi)
   \, ,\nonumber\\
   \delta\Gamma_{12}^{d\,\, ij}(Q)&=& 4\lambda^4f_\pi\delta^{ij} 
   {\mathcal I}(Q;m_\pi,m_\sigma)\, ,\nonumber\\
   \delta\Gamma_{12}^{e\,\, ij}(Q)&=& 6\lambda^4f_\pi\delta^{ij} 
   {\mathcal I}(Q;m_\sigma,m_\sigma)\, ,
   \label{vert12}  
\ee
where the function ${\mathcal I}$ is defined in the first of 
Eqs.~(\ref{funcIJ}) and the function ${\mathcal K}$ is defined by
\be
   {\mathcal K}(Q,P;m_i,m_j)\equiv T\sum_n\int\frac{d^3k}{(2\pi)^3}
   \frac{1}{K+m_i^2}\,\frac{1}{(K-Q)^2+m_i^2}\,\frac{1}{(K-Q-P)^2+m_j^2}\, . 
   \label{funcK}
\ee

The expression for $\delta\Gamma_{12}^{e\,\, ij}$, where the internal lines 
are sigma propagators, is exponentially suppressed in the kinematical regime 
considered and thus we drop it from the beginning. For the functions 
$\delta\Gamma_{12}^{b\,\, ij}$ to $\delta\Gamma_{12}^{d\,\, ij}$, notice also 
that in the kinematical regime where $m_\sigma \gg m_\pi,\, T$, we can 
{\it pinch} the internal sigma lines when considering the temperature 
dependent terms, given that the internal momentum is cut off by the 
temperature. In this approximation and after a suitable shift of the 
variable of integration and the analytical continuation in 
Eq.~(\ref{continuation}), it is easy to see that
\be
   {\mathcal K}(Q,P;m_\pi,m_\sigma)&\longrightarrow&
   -\frac{1}{m_\sigma^2}{\mathcal I}^t(Q;m_\pi,m_\pi)\, ,\nonumber\\
   {\mathcal K}(Q,P;m_\sigma,m_\pi)&\longrightarrow&
   \frac{1}{m_\sigma^4}{\mathcal J}(;m_\pi)\, ,\nonumber\\
   {\mathcal I}(Q;m_\pi,m_\sigma)&\longrightarrow&
   -\frac{1}{m_\sigma^2}{\mathcal J}(;m_\pi)\, .
   \label{limIK}
\ee
Therefore, adding up the expressions in Eq.~(\ref{vert12}) after the above
approximation, we have
\be
   \delta\Gamma_{12}^{ij}(Q) = 2\lambda^4f_\pi\delta^{ij}\left\{\left( 5
   -4\frac{\lambda^2 f_\pi^2}{m_\sigma^2}\right){\mathcal I}^t(Q;m_\pi,m_\pi) -
   \frac{2}{m_\sigma^2}\left(1 + 6\frac{\lambda^2f_\pi^2}{m_\sigma^2}\right)
   {\mathcal J}(;m_\pi)\right\}\, .
   \label{vert12ap}
\ee
As before, for the expression inside the curly brackets in
Eq.~(\ref{vert12ap}), we work to zeroth order in $m_\pi^2/m_\sigma^2$, thus, 
the effective one-sigma two-pion vertex to one-loop order can be written as
\be
   i\Gamma_{12}^{\star\, ij}(Q;P_1,P_2)&=&
   i\Gamma_{12}^{ij}+i\delta\Gamma_{12}^{ij}(Q)\nonumber\\
   &=&-2i\lambda^2 f_\pi\delta^{ij}
   \left[1 - 3\lambda^2{\mathcal I}^t(Q;m_\pi,m_\pi)\right]\, ,  
   \label{vert12tot}
\ee
using $i\Gamma_{12}^{ij}$ as given in Eq.~(\ref{rules}). We observe that
pinching the internal sigma lines and working at zeroth order in
$m_\pi^2/m_\sigma^2$ is effectively equivalent to consider that the
contributing diagrams are just those which can be made {\it topologically}
equivalent to the diagram represented by the function 
${\mathcal I}^t(Q;m_\pi,m_\pi)$. 

Finally, we consider the effective four-pion vertex. The diagrams contributing 
at one loop are depicted in Fig.~3. Pinching the internal sigma lines and 
working at zeroth order in $m_\pi^2/m_\sigma^2$, as before, one can check 
that the contributing diagrams are those coming from Figs.~3a to 3c, 
explicitly (hereafter, we omit the parametric dependence of the function 
${\mathcal I}^t$ on the mass $m_i$ for the sake of a clearer notation, except
when needed)
%%%%%%%%%%%%%%%%%%%%%%%%%%%%%%%%%%%%%%%%%%%%%%%%%%%%%%%%
\begin{figure}[h!] % fig 3
%\hspace{-2.5cm}
\centerline{\epsfig{file=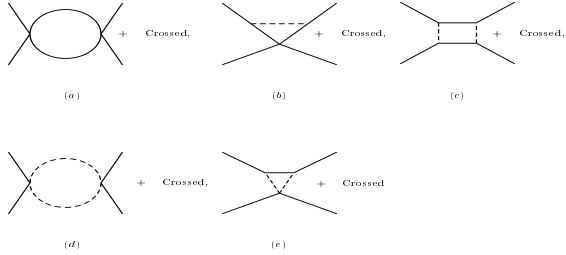,height=1.3in,width=3.0in}}
\vspace{1cm}
\caption{
Diagrams contributing to the four-pion vertex at one loop 
}
\end{figure}
%%%%%%%%%%%%%%%%%%%%%%%%%%%%%%%%%%%%%%%%%%%%%%%%%%%%%%%%
\be
   \delta\Gamma_{04}^{a\,\, ijkl}(;P_1,P_2,P_3,P_4)&=&2\lambda^4
   \left\{\left[ 7{\mathcal I}^t(P_1+P_2) + 
   2{\mathcal I}^t(P_1+P_3) + 2{\mathcal I}^t(P_1+P_4)
   \right]\delta^{ij}\delta^{kl}\right.\nonumber\\
   &+&\left[ 7{\mathcal I}^t(P_1+P_3) + 
   2{\mathcal I}^t(P_1+P_4) + 2{\mathcal I}^t(P_1+P_2)
   \right]\delta^{ik}\delta^{jl}\nonumber\\
   &+&\left.\left[ 7{\mathcal I}^t(P_1+P_4) + 
   2{\mathcal I}^t(P_1+P_2) + 2{\mathcal I}^t(P_1+P_3)
   \right]\delta^{il}\delta^{jk}\right\}\, ,\nonumber\\
   \delta\Gamma_{04}^{b\,\, ijkl}(;P_1,P_2,P_3,P_4)&=&
   4\lambda^4
   \left\{\left[{\mathcal I}^t(P_1+P_3) + {\mathcal I}^t(P_1+P_4)\right]
   \delta^{ij}\delta^{kl}\right.\, ,\nonumber\\
   &+&\left[{\mathcal I}^t(P_1+P_2) + {\mathcal I}^t(P_1+P_4)\right]
   \delta^{ik}\delta^{jl}\nonumber\\ 
   &+&\left.\left[{\mathcal I}^t(P_1+P_2) + {\mathcal I}^t(P_1+P_3)\right]
   \delta^{il}\delta^{jk}\right\}\nonumber\\
   \delta\Gamma_{04}^{c\,\, ijkl}(;P_1,P_2,P_3,P_4)&=& 
   -8\lambda^4
   \left[{\mathcal I}^t(P_1+P_2) + {\mathcal I}^t(P_1+P_3) + 
   {\mathcal I}^t(P_1+P_4)\right]\nonumber\\
   &&\left\{\delta^{ij}\delta^{kl} +
   \delta^{il}\delta^{jk} + \delta^{ik}\delta^{jl} \right\}\, .
   \label{vert4}
\ee
Adding up the expressions in Eq.~(\ref{vert4}) and using Eq.~(\ref{rules}), 
the effective four-pion vertex up to one loop can be written as
\be
   i\Gamma_{04}^{\star\ ijkl}(;P_1,P_2,P_3,P_4)&=&i\Gamma_{04}^{ijkl}+
   i\delta\Gamma_{04}^{ijkl}(;P_1,P_2,P_3,P_4)\nonumber\\ 
   &-&2i\lambda^2\left\{
   \left[1 - 3\lambda^2{\mathcal I}^t(P_1+P_2)\right]
   \delta^{ij}\delta^{kl}\right.\nonumber\\
   &+& 
   \left[1 - 3\lambda^2{\mathcal I}^t(P_1+P_3)\right]
   \delta^{ik}\delta^{jl}\nonumber\\
   &+& \left.
   \left[1 - 3\lambda^2{\mathcal I}^t(P_1+P_4)\right]\delta^{il}
   \delta^{jk}\right\}\, .
   \label{vert4tot}
\ee
It is easy to check that Eqs.~(\ref{newsigprop}), (\ref{vert12tot}) and 
(\ref{vert4tot}) satisfy the Ward identities in Eq.~(\ref{Ward}), this ensures
that the approximation scheme adopted respects chiral symmetry.

\section{Two-loop pion dispersion relation}\label{secIII}

We now use the above effective vertices and propagator to construct the
two-loop modification to the pion self-energy. The corresponding diagrams are 
shown in Fig.~4. Notice that diagrams 4$d$ and 4$e$ require knowledge of the 
effective, one-loop, two-sigma two-pion and three-sigma vertices, which we 
have not computed in this work. However, we can check that the leading 
contributions arise from diagrams 4$a$ to 4$c$. To see this, notice that 
the explicit expressions for diagrams 4$d$ and 4$e$ are equivalent to the
expressions for diagrams 4$a$ and 4$b$, respectively, upon the replacement of
the pion loop by a sigma loop. This results in expressions exponentially 
suppressed and thus, in our approximation, we should 
ignore them. The contributing terms are
%%%%%%%%%%%%%%%%%%%%%%%%%%%%%%%%%%%%%%%%%%%%%%%%%%%%%%%%
\begin{figure}[h!] % fig 4
%\hspace{-2.5cm}
\centerline{\epsfig{file=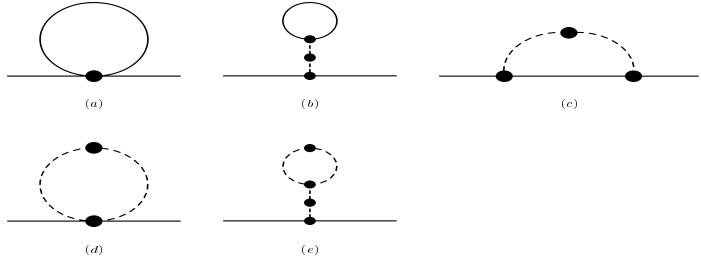,height=1.3in,width=3.0in}}
\vspace{1cm}
\caption{
 Diagrams contributing to the pion self-energy in the 
                 effective expansion. The heavy dots denote the effective
                 vertices and propagators.
}
\end{figure}
%%%%%%%%%%%%%%%%%%%%%%%%%%%%%%%%%%%%%%%%%%%%%%%%%%%%%%%%
\be
   \Pi^a_2(P)\ \delta^{ij}&=&\lambda^2
   \delta^{ij}\int_\beta\frac{d^4K}{(2\pi)^4}
   \frac{1}{K^2-m_\pi^2}\left\{5-9{\mathcal I}^t(0)-6\lambda^2{\mathcal I}^t
   (P+K)\right\}\nonumber\\
   \Pi^b_2(P)\ \delta^{ij}&=&-\lambda^2\delta^{ij}
   \left(\frac{6\lambda^2f_\pi^2}{m_\sigma^2}\right)\int_\beta
   \frac{d^4K}{(2\pi)^4}\frac{1}{K^2-m_\pi^2}
   \left\{\frac{[1-3\lambda^2{\mathcal I}^t(0)]^2}
   {[1-\frac{6\lambda^4f_\pi^2}{m_\sigma^2}{\mathcal I}^t(0)]}\right\}
   \nonumber\\
   \Pi^c_2(P)\ \delta^{ij}&=&-\lambda^2\delta^{ij}
   \left(\frac{4\lambda^2f_\pi^2}{m_\sigma^2}\right)\int_\beta
   \frac{d^4K}{(2\pi)^4}\frac{1}{K^2-m_\pi^2}
   \left\{\frac{[1-3\lambda^2{\mathcal I}^t(P+K)]^2}
   {[1-\frac{6\lambda^4f_\pi^2}{m_\sigma^2}
   {\mathcal I}^t(P+K)-\frac{(P+K)^2}{m_\sigma^2}]}\right\}\, ,
   \label{selfenergies}
\ee
where the subindex $\beta$ means that the integrals are to be computed at
finite temperature. We now expand the denominators in the second and third
of Eqs.~(\ref{selfenergies}), keeping only the leading order contribution
when considering $m_\pi$, $T$ and $P$ as small compared to $m_\sigma$. Adding
up the above three terms and to zeroth order in $m_\pi^2/m_\sigma^2$ where
\be
   \lambda^2\left(1-\frac{2\lambda^2f_\pi^2}{m_\sigma^2}\right)\approx
   \frac{m_\pi^2}{2f_\pi^2}\, , 
   \label{approx}
\ee
the pion self-energy can be written, in the imaginary-time formalism of TFT, as
\be
   \Pi_2(P)&=&\left(\frac{m_\pi^2}{2f_\pi^2}\right)
   T\sum_n\int\frac{d^3k}{(2\pi)^3}\frac{1}{K^2+m_\pi^2}
   \left\{5 + 2\left(\frac{P^2+K^2}{m_\pi^2}\right)\right.\nonumber\\
   &-&\left.\left(\frac{m_\pi^2}{2f_\pi^2}\right)
   \left[9{\mathcal I}^t(0) + 6{\mathcal I}^t(P+K)\right]\right\}\, ,
   \label{self}
\ee
with $K=(\omega_n,{\mathbf{k}})$ and $P=(\omega,{\mathbf{p}})$. The pion 
dispersion relation is thus obtained from the solution to
\be
   P^2 + m_\pi^2 +{\mbox R}{\mbox e}\Pi_2(P)=0\, ,
   \label{dispersion}
\ee
after renormalization and the analytical continuation 
$i\omega\rightarrow p_o + i\epsilon$. As anticipated, to compute 
${\mbox R}{\mbox e}\Pi_2(P)$ we require knowledge of the real part of 
${\mathcal I}^t$ only at $Q=0$ together with the full momentum dependence of 
the imaginary part of ${\mathcal I}^t$ (see Eq.~(\ref{reS}) below).  

Notice that Eq.~(\ref{self}) contains temperature-dependent infinities 
coming from the terms involving the function ${\mathcal I}^t$, as well as 
vacuum infinities. This is an usual feature of two-loop calculations at finite 
temperature where one always encounters temperature-dependent infinities in 
integrals involving only the bare terms of the original Lagrangian. However, 
as it turns out, these infinities are exactly canceled by the contribution 
from the integrals computed by using the counterterms that are necessary to 
introduce at one loop to carry the (vacuum) renormalization. The above was 
explicitly shown for the case of the self-energy in the $\phi^4$ theory by 
Kislinger and Morley~\cite{Kislinger} and for the sigma vacuum expectation 
value in the linear sigma model by Mohan~\cite{Mohan}. We make a brief sketch 
of the renormalization procedure for the pion self-energy at two loops in the 
appendix to show how this cancellation occurs and refer the reader to the 
cited works for details. From now on, we concentrate on the 
temperature-dependent terms.

Let us first look at the dispersion relation at leading order. After analytical
continuation all the terms are real. The integrals involved are 
\be
   T\sum_n\int\frac{d^3k}{(2\pi)^3}\frac{1}{K^2+m_\pi^2}&\rightarrow&
   \frac{1}{2\pi^2}\int_0^\infty\frac{dk\ k^2}{E_k}f(E_k)\nonumber\\
   &\equiv&\frac{m_\pi^2}{2\pi^2}\ g(m_\pi/T)\nonumber\\
   T\sum_n\int\frac{d^3k}{(2\pi)^3}\frac{K^2}{K^2+m_\pi^2}&\rightarrow&
   -\ m_\pi^2\ \left(\frac{m_\pi^2}{2\pi^2}\right)\ g(m_\pi/T)\, ,
   \label{g}
\ee 
where $g$ is the dimensionless function defined in the second of 
Eqs.~(\ref{temprefuncs}) and the arrows indicate only the temperature 
dependence of the expressions. Thus, the dispersion relation results from
\be
   \left[1 + 2\ \xi\ g(m_\pi/T)\right](p_0^2-p^2) -
   \left[1 + 3\ \xi\ g(m_\pi/T)\right]m_\pi^2 = 0\, ,
   \label{dis1a}
\ee
with $\xi=m_\pi^2/4\pi^2f_\pi^2\ll 1$. For $T\sim m_\pi$, $g(m_\pi/T)\sim1$,  
therefore, at leading order and in the kinematical regime that we are 
considering, Eq.~(\ref{dis1a}) can be written as 
\be
   p_0^2 = p^2 + m_\pi^2\left[1 + \xi\ g(m_\pi/T)\right]\, ,
   \label{dis1b}
\ee
which coincides with the result obtained from ChPT~\cite{Gasser}. 

We now look at the next to leading order terms in Eq.~(\ref{self}). The first
of these is purely real and represents a constant, second order shift to the
pion mass squared 
\be
   -\ 9\left(\frac{m_\pi^2}{2f_\pi^2}\right)^2
   T\sum_n\int\frac{d^3k}{(2\pi)^3}\frac{{\mathcal I}^t(0)}{K^2+m_\pi^2}
   \rightarrow
   \frac{9}{2}\ \xi^2\
   g(m_\pi/T)\ h(m_\pi/T)\ m_\pi^2\, ,
   \label{secordmass}
\ee
where $h$ is the dimensionless function defined in the first of 
Eqs.~(\ref{temprefuncs}). The remaining term in Eq.~(\ref{self}) shows a 
non-trivial dependence on $P$. It involves the function ${\mathcal S}$ defined 
by
\be
   {\mathcal S}(P)\equiv T\sum_n\int\frac{d^3k}{(2\pi)^3}
   \frac{{\mathcal I}^t(P+K)}{K^2+m_\pi^2}\, .
   \label{s}
\ee
The sum is performed by resorting to the spectral representation of 
${\mathcal I}^t$ and $(K^2+m_\pi^2)^{-1}$. Thus, the real part of the 
retarded version of ${\mathcal S}$, after analytical continuation is
\be
   {\mbox R}{\mbox e}{\mathcal S}^r(p_0,p)&\equiv&\frac{1}{2}
   \left[{\mathcal S}(i\omega\rightarrow p_0+i\epsilon ,p) + 
   {\mathcal S}(i\omega\rightarrow p_0-i\epsilon ,p)\right]\nonumber\\
   &=&\ -\ {\mathcal P}\int\frac{d^3k}{(2\pi)^3}
   \int_{-\infty}^{\infty}\frac{dk_0}{2\pi}\int_{-\infty}^{\infty}
   \frac{dk_0'}{2\pi}\left[1+f(k_0)+f(k_0')\right]\nonumber\\
   &&
   \frac{2\pi\ \varepsilon(k_0')\ 
   \delta[{k_0'}^2-({\mathbf{k}}-{\mathbf{p}})^2
   -m_\pi^2]\ 2\ {\mbox I}{\mbox m}{\mathcal I}^t(k_0,k)}{p_0-k_0-k_0'}\, ,
   \label{reS}
\ee
where ${\mathcal P}$ stands for the principal part of the integral. 

Including all the terms, the dispersion relation up to next to leading order
in the parameter $\xi$, for $T\sim m_\pi$ and in the small momentum region is 
obtained as the solution to
\be
   p_0^2=p^2 +
   \left\{1 + \xi g(m_\pi/T) + 
   \frac{\xi^2}{2}g(m_\pi/T)\left[9h(m_\pi/T)-4g(m_\pi/T)\right]
   \right\}m_\pi^2 + \xi^2\tilde{\mathcal S}(p_0,p)\, ,
   \label{disp}
\ee
where $\tilde{\mathcal S}$ is defined by
\be
   \tilde{\mathcal S}(p_0,p)= -(24\pi^4){\mbox R}{\mbox e}{\mathcal S}^r
   (p_0,p)\, .
   \label{tilS}
\ee
%%%%%%%%%%%%%%%%%%%%%%%%%%%%%%%%%%%%%%%%%%%%%%%%%%%%%%%%
\begin{figure}[h!] % fig 5
%\hspace{-2.5cm}
\centerline{\epsfig{file=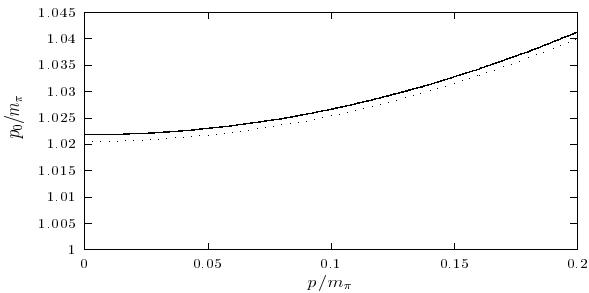,height=1.5in,width=3.0in}}
\vspace{1cm}
\caption{
Pion dispersion relation obtained as the solution to
                 Eq.~(\ref{disp}) for $T=m_\pi$ (upper curve). Shown is also
                 the dispersion relation obtained by ignoring the term
                 $\tilde{\mathcal S}(p_0,p)$ (lower curve).
}
\end{figure}
%%%%%%%%%%%%%%%%%%%%%%%%%%%%%%%%%%%%%%%%%%%%%%%%%%%%%%%%%%%
%%%%%%%%%%%%%%%%%%%%%%%%%%%%%%%%%%%%%%%%%%%%%%%%%%%%%%%%
\begin{figure}[h!] % fig 6
%\hspace{-2.5cm}
\centerline{\epsfig{file=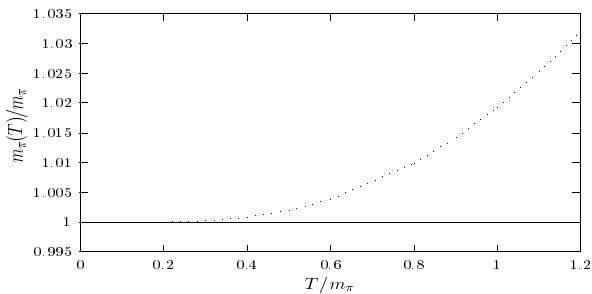,height=1.5in,width=3.0in}}
\vspace{1cm}
\caption{
Temperature dependence of $m_\pi$ to second order in the
                 parameter $\xi$. For comparison, the vacuum value is shown as 
                 a horizontal line. 
}
\end{figure}
%%%%%%%%%%%%%%%%%%%%%%%%%%%%%%%%%%%%%%%%%%%%%%%%%%%%%%%%%%%
Figure~5 shows the dispersion relation obtained from Eq.~(\ref{disp})
for $T=m_\pi$ where we also display the solution without the term 
$\xi^2\tilde{\mathcal S}(p_0,p)$. Inclusion of this last term does not alter 
the shape of the curve in this kinematical regime. Since
$\tilde{\mathcal S}(p_0,p)$ is non-vanishing for $p=0$, it also contributes to
the perturbative increase of the pion mass. Figure~6 shows the temperature
dependence of $m_\pi$ obtained as the solution to Eq.~(\ref{disp}) in the
limit when $p$ goes to zero. Notice that the second order correction in the
parameter $\xi$ has the same sign as the first order correction. This result
is opposite to the temperature behavior of the pion mass found in 
Refs.~\cite{Schenk,Toublan}.  

\section{Pion damping}\label{secIV}

Damping of pions traveling in the pion medium is due to two possible kinds of 
processes: formation of resonances and pion elastic scattering. At 
temperatures on the order of the pion mass, pion elastic scattering is 
subdominant since, given their Goldston-boson nature, scattering among pions 
nearly disappears as the relative momentum approaches zero. In the linear sigma
model, this feature is explicitly realized, for the kinematical regime 
discussed in this work, by the description of pion elastic scattering as a
second order process in the small parameter $\xi$. Scattering of a pion
off a given phase space element as a consequence of an elastic
process is described by means of the imaginary part of $\Pi_2$ given in 
Eq.~(\ref{self}). This can be understood by observing that working in the 
approximation where $m_\sigma \gg m_\pi,\, T$ the contributing diagrams to 
$\Pi_2$ can be thought of as made exclusively of pion lines, with effective 
one-loop vertices that involve also only pion lines. Thus, the rate obtained 
by cutting such diagrams describes a processes in which pions scatter from 
each other and keep being pions. We will discuss elastic pion scattering 
later on in this section. First, we look at resonance formation.

%%%%%%%%%%%%%%%%%%%%%%%%%%%%%%%%%%%%%%%%%%%%%%%%%%%%%%%%
\begin{figure}[h!] % fig 7
%\hspace{-2.5cm}
%\centerline{\epsfig{file=pionprop3fig7.eps,height=7.0in,width=6.0in}}
\centerline{\psfig{file=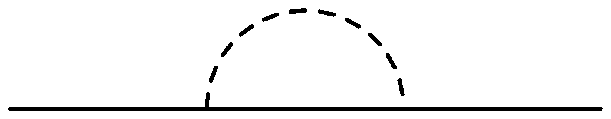,height=1.5in,width=3.5in}}
\vspace{-1.3cm}
%\hspace{-2.5cm}
\caption{
Relevant diagram to compute pion damping by the formation of
sigma resonances.
}
\end{figure}
%%%%%%%%%%%%%%%%%%%%%%%%%%%%%%%%%%%%%%%%%%%%%%%%%%%%%%%%%%%
In the linear sigma model with only meson degrees of freedom, pions can
disappear from the medium by forming a sigma resonance. This process is the
inverse of the one describing sigma decay (except that the former does not
happen at T=0) and thus both rates are intimately connected. The relevant 
diagram to compute resonance formation is depicted in
Fig.~7 and its explicit expression is
\be
   \Pi_1(P)=4\lambda^4f_\pi^2{\mathcal I}(P;m_\pi,m_\sigma)\, ,
   \label{self1}
\ee
where ${\mathcal I}$ is the function defined in Eq.~(\ref{funcIJ}) that we
rewrite here in terms of the quantities $E_\pi=\sqrt{k^2+m_\pi^2}$ and
$E_\sigma=\sqrt{({\mathbf{k}}-{\mathbf{p}})^2+m_\sigma^2}$ as
\be
  {\mathcal I}(P;m_\pi,m_\sigma)=T\sum_n\int\frac{d^3k}{(2\pi)^3}\,\frac{1}
  {\omega_n^2+E_\pi^2}\,\frac{1}{(\omega_n-\omega)^2+E_\sigma^2}\, .
  \label{newI}
\ee
The damping rate is obtained in terms of the imaginary part of retarded 
version, ${\mathcal I}^r$ of ${\mathcal I}(P;m_\pi,m_\sigma)$, given by
\be
   {\mbox I}{\mbox m}{\mathcal I}^r(p_0,p;m_\pi,m_\sigma)
   &=&-\pi\int\frac{d^3k}{(2\pi)^3}
   \frac{1}{4E_\pi E_\sigma}\nonumber\\
   &&\left\{[1+f(|E_\pi|)+f(|E_\sigma|)]
   [\delta(p_0 - E_\pi - E_\sigma) -
   \delta(p_0 + E_\pi + E_\sigma)]\right.\nonumber\\ 
   &-&\left.[f(|E_\pi|)-f(|E_\sigma|)]
   [\delta(p_0 - E_\pi + E_\sigma) - \delta(p_0 + E_\pi - E_\sigma)]
   \right\}\, ,
   \label{im1}
\ee
after the analytical continuation $i\omega\rightarrow p_o+i\epsilon$. The
function $f$ above is the Bose-Einstein distribution and its arguments here are
the absolute values of the corresponding particle energies. The integration in
Eq.~(\ref{im1}) can be performed analytically and the result is
\be
   {\mbox I}{\mbox m}{\mathcal I}^r(p_0,p;m_\pi,m_\sigma)
   = -\frac{T}{16\pi p}\left\{\ln\left(\frac{1-e^{-E_+(p_0,p)/T}}
   {1-e^{-E_-(p_0,p)/T}}\right) - \ln\left(
   \frac{1-e^{-(E_+(p_0,p) + p_0)/T}}
   {1-e^{-(E_-(p_0,p) + p_0)/T}}\right) \right\}\, ,
   \label{imana}
\ee
where the functions $E_\pm$ are defined by
\be
   E_\pm(p_0,p)=b\left(\frac{p_0\pm a'(P^2)p}{2}\right)\, ,
   \label{es}
\ee
with
\be
   a'(P^2)&=&\sqrt{1-\frac{4m_\pi^2}{b^2P^2}}\, ,\nonumber\\
   b&=&\frac{m_\sigma^2-m_\pi^2}{m_\pi^2} - 1\, .
   \label{ab}
\ee
The decay rate $\Gamma^>_1$ is given in terms of ${\mbox I}{\mbox m}\Pi_1$ by
\be
   \Gamma^>_1(p)=-\frac{e^{p_0/T}}{(e^{p_0/T}-1)}\,
   \frac{{\mbox I}{\mbox m}\Pi_1(p_0,p)}{p_0}
   \left|_{p_0=\sqrt{p^2+m_\pi^2}}\right.\, .
   \label{decayr}
\ee
${\mbox I}{\mbox m}\Pi_1$ contains both creation and decay rates and the
factor $e^{p_0/T}/(e^{p_0/T}-1)$ in Eq.~(\ref{decayr}) eliminates the piece
describing pion creation.

We can now compute the mean free path $\lambda$ for a pion traveling in the 
medium before forming a sigma resonance. This is given in terms of $\Gamma^>_1$
by~\cite{LeBellac}
\be
   \lambda &=& \frac{v}{\Gamma^>_1(p)}\nonumber\\
   &=&-\frac{(e^{p_0/T}-1)}{e^{p_0/T}}\,
   \frac{p}{{\mbox I}{\mbox m}\Pi_1(p_0,p)}
   \left|_{p_0=\sqrt{p^2+m_\pi^2}}\right.\, ,
   \label{freep}
\ee
where $v=p/p_0$ is the magnitude of the pion's velocity. Figure~8 shows plots 
of $\lambda$ for three different temperatures and a value of $m_\sigma = 600$ 
MeV as a function of the pion momentum. $\lambda$ reaches a maximum for
$p\sim 0.4 m_\pi$. Notice that the position of the maximum is approximately
independent of $T$. Except for the location of the maximum, the behavior of
the curves is both in qualitative and quantitative agreement with the 
corresponding result obtained in ChPT (see Fig. 2 Ref.~\cite{Goity}).

%%%%%%%%%%%%%%%%%%%%%%%%%%%%%%%%%%%%%%%%%%%%%%%%%%%%%%%%
\begin{figure}[h!] % fig 8
%\hspace{-2.5cm}
\centerline{\epsfig{file=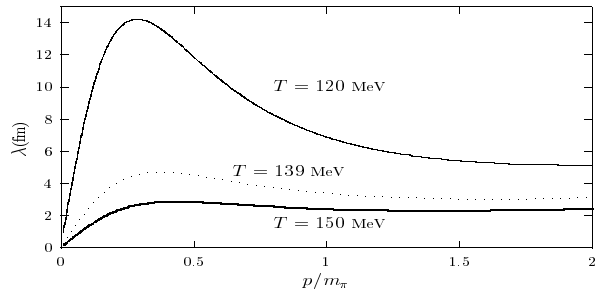,height=1.5in,width=3.0in}}
\vspace{1cm}
\caption{
Mean free path as a function of momentum for a pion to travel
                 the medium before forming a sigma resonance for
                 $m_\sigma=600$ MeV.
}
\end{figure}
%%%%%%%%%%%%%%%%%%%%%%%%%%%%%%%%%%%%%%%%%%%%%%%%%%%%%%%%%%%
%%%%%%%%%%%%%%%%%%%%%%%%%%%%%%%%%%%%%%%%%%%%%%%%%%%%%%%%
\begin{figure}[h!] % fig 9
%\hspace{-2.5cm}
\centerline{\epsfig{file=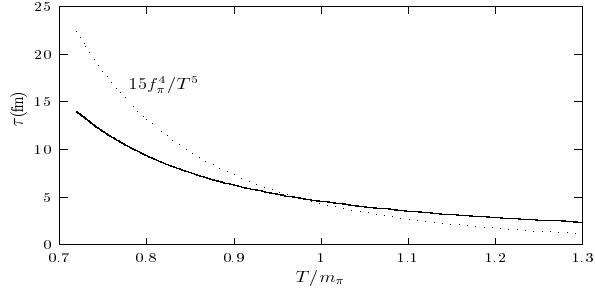,height=1.5in,width=3.0in}}
\vspace{1cm}
\caption{
Mean collision time as a function of temperature for a pion 
                 to travel the medium before forming a sigma resonance. Shown
                 is also the function $15f_\pi^4/T^5$.
}
\end{figure}
%%%%%%%%%%%%%%%%%%%%%%%%%%%%%%%%%%%%%%%%%%%%%%%%%%%%%%%%%%%
We can also compute the mean collision time $\tau$ between pions before 
forming a sigma resonance. This quantity is defined as the inverse of the 
average decay rate
\be
   \tau^{-1}&\equiv&\langle\Gamma^>_1\rangle\nonumber\\
   &=&\int\frac{d^3p}{(2\pi)^3}f\left(\sqrt{p^2+m_\pi^2}\,\right)\,\Gamma^>_1
   \left/
   \int\frac{d^3p}{(2\pi)^3}f\left(\sqrt{p^2+m_\pi^2}\,\right)\right.\, ,
   \label{average}
\ee
where $f$ is the Bose-Einstein distribution. Figure~9 shows the behavior of
$\tau$ as a function of the temperature for $m_\sigma=600$ MeV. Shown is also
the quantity $15f_\pi^4/T^5$. We see that $\tau$ does not behave exactly as the
inverse fifth power of the temperature~\cite{Shuryak2}, though the curves are
pretty close to each other in the temperature range considered. The pion
thermal width has also been studied in Refs.~\cite{Dominguez}~\cite{Leutwyler}.

Finally, let us discuss pion elastic scattering. The total rate, from where
the cross section for elastic pion scattering can be obtained, is found in 
terms of the imaginary part of the two-loop pion self-energy $\Pi_2$ given in
Eq.~(\ref{self})
\be
   {\mbox I}{\mbox m}\Pi_2(p_0,p)=-24\pi^4\xi^2
   {\mbox I}{\mbox m}{\mathcal S}^r(p_o,p)\, ,
   \label{imaS}
\ee
where ${\mbox I}{\mbox m}{\mathcal S}^r$ is the imaginary part of the retarded 
version of the function ${\mathcal S}$ defined in Eq.~(\ref{s}),
explicity~\cite{LeBellac} 
\be
   {\mbox I}{\mbox m}{\mathcal S}^r(p_0,p)&\equiv&\frac{1}{2i}
   \left[{\mathcal S}(i\omega\rightarrow p_0+i\epsilon ,p) - 
   {\mathcal S}(i\omega\rightarrow p_0-i\epsilon ,p)\right]\nonumber\\
   &=&\pi\left(e^{p_0/T}-1\right)\int\frac{d^3k}{(2\pi)^3}
   \int_{-\infty}^{\infty}\frac{dk_0}{2\pi}\int_{-\infty}^{\infty}
   \frac{dk_0'}{2\pi}f(k_0)f(k_0')\delta(p_0-k_0-k_0')\nonumber\\
   &&
   2\pi\ \varepsilon(k_0')\ 
   \delta[{k_0'}^2-({\mathbf{k}}-{\mathbf{p}})^2
   -m_\pi^2]\ 2\ {\mbox I}{\mbox m}{\mathcal I}^t(k_0,k)\, .
   \label{imaSexpl}
\ee
%%%%%%%%%%%%%%%%%%%%%%%%%%%%%%%%%%%%%%%%%%%%%%%%%%%%%%%%
\begin{figure}[h!] % fig10 
%\hspace{-2.5cm}
\centerline{\epsfig{file=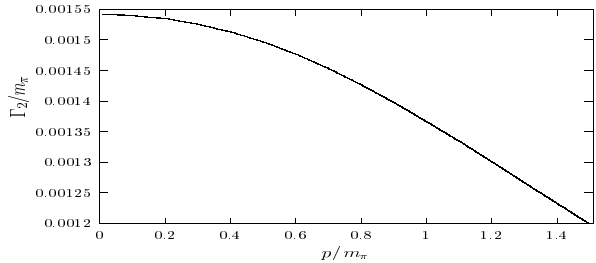,height=1.35in,width=3.0in}}
\vspace{1cm}
\caption{
Total pion elastic scattering rate $\Gamma_2$ for $T=m_\pi$
                  as a function of momentum. Notice that this rate is about 
                  two orders of magnitude smaller than $\Gamma_1$, given that 
                  it is proportional to $\xi^2$.
}
\end{figure}
%%%%%%%%%%%%%%%%%%%%%%%%%%%%%%%%%%%%%%%%%%%%%%%%%%%%%%%%%%%
%%%%%%%%%%%%%%%%%%%%%%%%%%%%%%%%%%%%%%%%%%%%%%%%%%%%%%%%
\begin{figure}[h!] % fig11 
%\hspace{-2.5cm}
\centerline{\epsfig{file=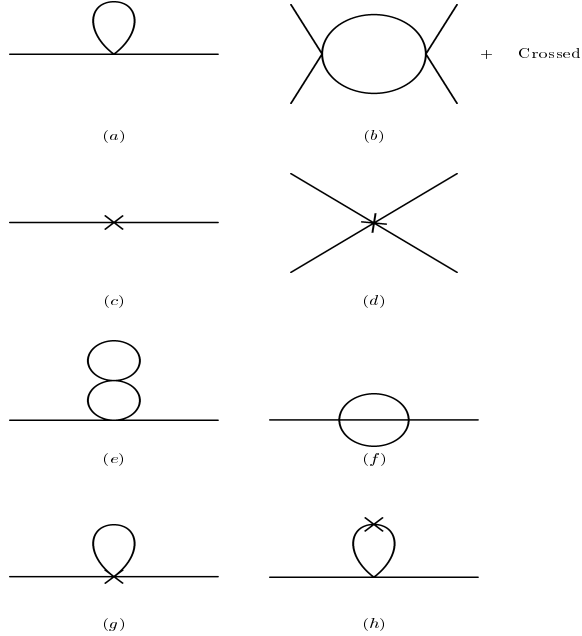,height=3.3in,width=3.0in}}
\vspace{1cm}
\caption{
 One and two-loop diagrams and counterterms for the
                  computation of the pion-self energy described by the
                  effective Lagrangian in Eq.~(\ref{effLag}). The
                  temperature-dependent infinities are exactly canceled.
}
\end{figure}
%%%%%%%%%%%%%%%%%%%%%%%%%%%%%%%%%%%%%%%%%%%%%%%%%%%%%%%%%%%
The total reaction rate is given by
\be
   \Gamma_2(p_o,p)=-\,\frac{1}{p_0}{\mbox I}{\mbox m}\Pi_2(p_0,p)\, ,
   \label{rate2}
\ee
where $p_0$ should be taken as the solution to Eq.~(\ref{dis1b}). The 
integration in Eq.~(\ref{imaSexpl}) can only be performed numerically. 
Figure~10 shows a plot of $\Gamma_2$ as a function of $p$ for $T=m_\pi$. As 
anticipated, this rate is about two orders of magnitude smaller than 
$\Gamma_1$, given that it is proportional to $\xi^2$. 

\section{Summary and conclusions}\label{secV}

In conclusion, working in the linear sigma model at finite temperature, we 
have found effective one-loop sigma propagator and one-sigma two-pion and 
four-pion vertices satisfying the ChWI and therefore respecting chiral 
symmetry, when maintaining only the zeroth order terms in an expansion in the 
parameter $m_\pi^2/m_\sigma^2$. We have used these objects to compute the
two-loop order correction to the pion propagator in a pion medium for small 
momentum and for $T\sim m_\pi$ and showed that the linear sigma model yields 
the same result as ChPT at leading order in the parameter 
$\xi=m_\pi^2/4\pi^2f_\pi^2$. This result was to be expected since the 
kinematical regime we consider is that where the temperature, the pion 
momentum and the pion mass are treated as small quantities, such as in the 
case of ChPT. The existence of the parameter $\xi$ allows for a controlled
perturbative calculation of physical quantities in the above mentioned
kinematical regime, such as the thermal modification of the pion mass. We have
shown that, contrary to the result in Refs.~{\cite{Schenk,Toublan}, the 
two-loop order correction to the pion mass is proportional to $\xi^2$ and that 
this correction is of the same sign as the one-loop correction. The shape of 
the dispersion curve is not significantly altered in the kinematical regime
considered. We should also note that our result is more general than the
one obtained in Ref.~\cite{Pisarski}, where the computation of the two-loop
pion dispersion relation is carried from the onset in the weak coupling
limit, that is to say, for $\lambda^2 \ll 1$. In our approach, the first
step can be thought of as valid for arbitrary values of $\lambda^2$, this
is why the effective one-loop vertices and propagator found depend
explicitly on $\lambda^2$. However, when it comes to the computation of a
physical quantity such as the pion dispersion relation, we have shown 
(see Eq.~(\ref{approx})) that the contribution from $\lambda^2$ enters the
result always in the combination 
\be
   \lambda^2\left( 1 - \frac{2\lambda^2 f_\pi^2}{m_\sigma^2}\right)&=&
   \lambda^2\left( 1-1+\frac{m_\pi^2}{m_\sigma^2} \right)\nonumber\\
   &=& \lambda^2\frac{m_\pi^2}{m_\sigma^2}\nonumber\\
   &=& \frac{m_\pi^2}{2f_\pi^2} + {\mathcal{O}}(m_\pi^2/m_\sigma^2)
   \label{explan}
\ee
showing that for $m_\sigma > m_\pi$, the leading contribution is
independent of $m_\sigma$.

We have also studied pion damping both in the elastic and inelastic
channels. We have computed the momentum dependence of the mean free path for 
a pion before forming a sigma resonance and found that the shape of the curves
coincide both quantitatively and qualitatively with the result obtained from
ChPT when using a value of $m_\sigma=600$ MeV. The mean collision time of a 
pion before forming a sigma resonance as a function of $T$ is however close
but not quite proportional to the inverse fifth power of the temperature. It 
remains to include the effects of a possible high nucleonic density in the same
kinematical regime. 

Regarding the applicability of the linear sigma model as a tool to study low
temperature QCD, let us stress that at such energies, any theory
whose aim is to describe such degrees of freedom is an
effective one. The essential ingredient, common to all of them is
chiral symmetry, that is to say, the consequences arising from the
smallness of the pion mass. Therefore, as long as chiral symmetry is
respected and the temperature is small enough (on the order of the pion
mass) so as to not excite higher resonances, all such theories will
yield equivalent results. We have shown explicitly that this is the case
since, contrary to common belief, the one-loop computation of the pion 
dispersion curve yields the same result as the one obtained by using ChPT, 
provided we work in the same kinematical regime.

We should also mention that for temperatures on the order of 150 MeV, it 
might be important to include the effects introduced by other degrees of
freedom, such as the $\rho$ meson. In this sense, the quantities computed
in this work at such temperatures must be regarded as examples of the
results obtained by considering only the lowest lying meson degrees of
freedom and by no means as the final answer to the pion propagation
properties.

Finally, it is interesting to note that the formalism thus 
developed could be employed to explore the behavior of the pion dispersion 
curve in the large momentum region where, in principle, the lowest order ChPT 
Lagrangian cannot be used. These issues will be treated in a follow up 
work~\cite{progress}.

\section*{Acknowledgments}

The authors acknowledge useful conversations with M. Napsuciale. Support for 
this work has been received in part by CONACyT M\'exico under grant number 
32279-E.

\section*{Appendix: Two-loop renormalization at finite temperature}

The purpose of this appendix is to show how the cancellation of
the temperature-dependent infinities appearing in two loop computations is
achieved. For this matter, we concentrate on the two-loop pion self-energy
given in Eq.~(\ref{self}). Notice that the first and third terms can be nicely 
combined to give a numerical factor of $3$. Moreover, we carry the
renormalization procedure at $P=0$ where the approximation used renders the
result exact. Thus, Eq.~(\ref{self}) can be written as
\be
   \Pi_2(P)=\left(\frac{m_\pi^2}{2f_\pi^2}\right)
   T\sum_n\int\frac{d^3k}{(2\pi)^3}\frac{1}{K^2+m_\pi^2}
   \left\{3-\left(\frac{m_\pi^2}{2f_\pi^2}\right)
   \left[9{\mathcal I}^t(0) + 6{\mathcal I}^t(P+K)\right]\right\}\, .
   \label{selfmod}
\ee
The two-loop self-energy expressed in Eq.~(\ref{selfmod}) can be formally 
obtained by means of the effective Lagrangian
\be
   {\mathcal L}=\frac{1}{2}\left(\partial_\mu{\mathbf{\phi}}\right)^2
   -\frac{1}{2}m_\pi^2{\mathbf{\phi}}^2 -\frac{\alpha}{4}\left(
   {\mathbf{\phi}}^2\right)^2\, ,
   \label{effLag}
\ee
where $\alpha=6(m_\pi^2/2f_\pi^2)$. $m_\pi$ and $f_\pi$ are taken as the
renormalized values of the pion mass and decay constant, respectively. The 
factor $6$ comes from considering the interaction of like-isospin pions in 
the vertex
\be
   i\Gamma_4^{ijkl}=-2i\left(\frac{m_\pi^2}{2f_\pi^2}\right)
   \left(\delta^{ij}\delta^{kl}+\delta^{ik}\delta^{jl}+\delta^{il}\delta^{jk}
   \right)\, .
   \label{vertmod}
\ee
At one-loop, there are two (temperature-independent) infinities coming from
diagrams $a$ and $b$ in Fig.~11. Using the effective Lagrangian in
Eq.~(\ref{effLag}) and dimensional regularization, the counterterms needed to 
cancel these infinities, represented by diagrams $c$ and $d$ in Fig.~11, are
given explicitly by
\be
   \delta m_\pi^2 &=& 3\left(\frac{m_\pi^2}{2f_\pi^2}\right)
   \frac{m_\pi^2}{16\pi^2\epsilon}\, ,\nonumber\\
   \delta\alpha &=& 3\times36\left(\frac{m_\pi^2}{2f_\pi^2}\right)^2
   \frac{\mu^\epsilon}{16\pi^2\epsilon}\, ,
   \label{counter}
\ee
where $\epsilon=d-4$, $d$ being the number of dimensions, and the factor $3$
in the second of Eqs.~(\ref{counter}) takes care of the three crossed channels
represented by Fig.~11$b$, since the infinity in each of these is the same.
$\mu$ is the usual mass parameter introduced in dimensional regularization.

Going to two loops, temperature-dependent infinities appear. Diagram
$e$ in Fig.~11, which explicit expression is given by the second term in
Eq.~(\ref{selfmod}) gives rise to two infinite, temperature-dependent terms, 
coming from the product of the vacuum infinity in ${\mathcal I}^t(0)$ times the
temperature-dependent term in the integration and the vacuum infinity in the
integration times the temperature-dependent piece in ${\mathcal I}^t(0)$.
These are explicitly
\be
   {\mathcal Y}_1 &=& -9\times 2\left(\frac{m_\pi^2}{2f_\pi^2}\right)^2
   \frac{i\mu^\epsilon}{16\pi^2\epsilon}\left(\frac{m_\pi^2}{2\pi^2}\right)
   g(m_\pi/T)\, ,\nonumber\\
   {\mathcal Y}_2 &=& -9\left(\frac{m_\pi^2}{2f_\pi^2}\right)^2
   \frac{im_\pi^2}{16\pi^2\epsilon}\,{\mathcal I}^t_T(0)\, ,
   \label{inf12}
\ee
where $g(m_\pi/T)$ is the dimensionless function defined in the second of 
Eqs.~(\ref{temprefuncs}) and ${\mathcal I}^t_T(0)$ represents the
temperature-dependent part of the function ${\mathcal I}^t(0)$. 

Diagram $f$ in Fig.~11, which explicit expression corresponds to the third
term in Eq.~(\ref{selfmod}), gives rise to the temperature-dependent infinity
\be
   {\mathcal Y}_3 = -36\left(\frac{m_\pi^2}{2f_\pi^2}\right)^2
   \frac{i\mu^\epsilon}{16\pi^2\epsilon}\left(\frac{m_\pi^2}{2\pi^2}\right)
   g(m_\pi/T)\, .
   \label{inf3}
\ee
To complete the calculation, we need to compute the integrals involving the
one-loop counterterms. Diagram $g$ in Fig.~11 gives the explicit
temperature-dependent infinity
\be
   {\mathcal Y}_4 = \frac{3}{2}\times
   36\left(\frac{m_\pi^2}{2f_\pi^2}\right)^2
   \frac{i\mu^\epsilon}{16\pi^2\epsilon}\left(\frac{m_\pi^2}{2\pi^2}\right)
   g(m_\pi/T)\, ,
   \label{inf4}
\ee
whereas diagram $h$ in Fig.~11 gives
\be
   {\mathcal Y}_5 = 9\left(\frac{m_\pi^2}{2f_\pi^2}\right)^2
   \frac{im_\pi^2}{16\pi^2\epsilon}\,{\mathcal I}^t_T(0)\, .
   \label{inf5}
\ee
We can now see that ${\mathcal Y}_5$ in Eq.~(\ref{inf5}) cancels
${\mathcal Y}_2$ in the second of Eqs.~(\ref{inf12}) and that 
${\mathcal Y}_1,\,{\mathcal Y}_3$ and ${\mathcal Y}_4$ in 
Eqs.~(\ref{inf12}),~(\ref{inf3})~and~(\ref{inf4}), cancel among
themselves. Therefore, the result at two loops is free form
temperature-dependent infinities. Similar arguments apply to other Green's
functions, see Refs.\cite{Mohan,Kislinger} for additional details.

\end{document}